\documentclass{article}

\usepackage{PRIMEarxiv}
\usepackage[round]{natbib}
\usepackage{amsmath} 
\usepackage{amssymb}
\usepackage[tableposition=top]{caption}
\usepackage{subcaption}

\usepackage[utf8]{inputenc} 
\usepackage[T1]{fontenc}    
\usepackage{hyperref}       
\usepackage{url}            
\usepackage{booktabs}       
\usepackage{amsfonts}       
\usepackage{nicefrac}       
\usepackage{microtype}      
\usepackage{lipsum}
\usepackage{fancyhdr}       
\usepackage{graphicx}       

\title{A Multimodal Symphony: Integrating Taste and Sound through Generative AI
}

\author{
  Matteo Spanio\\
  Centro di Sonologia Computazionale (CSC) \\
  Department of Information Engineering \\
  University of Padova \\
  Padova, Italy\\
  \texttt{spanio@dei.unipd.it} \\
   \And
  Massimiliano Zampini\\
  Center for Mind/Brain Sciences (CIMeC) \\
  University of Trento \\
  Rovereto, Italy\\
  \texttt{massimiliano.zampini@unitn.it} \\
   \And
  Antonio Rodà \\
  Centro di Sonologia Computazionale (CSC) \\
  Department of Information Engineering \\
  University of Padova \\
  Padova, Italy\\
  \texttt{roda@dei.unipd.it} \\
   \And
  Franco Pierucci\\
  SoundFood s.r.l. \\
  Terni, Italy\\
  \texttt{franco.pierucci@soundfood.it}\\
}

\begin{document}
\maketitle

\begin{abstract}
In recent decades, neuroscientific and psychological research has traced direct relationships between taste and auditory perceptions. This article explores multimodal generative models capable of converting taste information into music, building on this foundational research. We provide a brief review of the state of the art in this field, highlighting key findings and methodologies. 
We present an experiment in which a fine-tuned version of a generative music model (MusicGEN) is used to generate music based on detailed taste descriptions provided for each musical piece. The results are promising: according the participants' ($n=111$) evaluation, the fine-tuned model produces music that more coherently reflects the input taste descriptions compared to the non-fine-tuned model. This study represents a significant step towards understanding and developing embodied interactions between AI, sound, and taste, opening new possibilities in the field of generative AI. We release our dataset, code and pre-trained model at: \url{https://osf.io/xs5jy/}.
\end{abstract}

\keywords{Generative AI \and Crossmodal correspondences \and Taste \and Audition \and Music}

\section{Introduction}\label{sec:intro}
\begin{figure}[h!]
\begin{center}
\includegraphics[width=10cm]{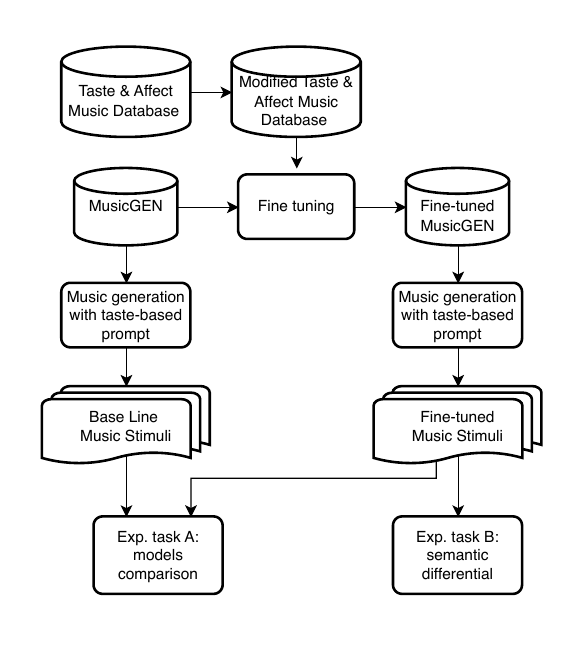}
\end{center}
\caption{Experimental pipeline for evaluating taste-based music generation. The Taste \& Affect Music Database is modified and used to fine-tune MusicGEN, resulting in a Fine-tuned MusicGEN model. Both the base and fine-tuned models generate music using taste-based prompts. The generated stimuli are then evaluated through two experimental tasks: (A) a comparison between baseline and fine-tuned models, and (B) a semantic differential analysis of the fine-tuned stimuli.}\label{fig:pipeline}
\end{figure}

Over recent years, the rapid evolution and progress of generative models have opened new possibilities in manipulating images, audio, and text, both independently and in a multimodal context. These AI advancements have ignited considerable debate about the essence of these human-engineered ``intelligences''. Critics have termed large language models (LLMs) as ``statistical parrots'' \citep{bender_stochastic_parrots_2021} due to their reliance on data. However, others view them as advanced tools capable of emulating and exploring the intricate structures of the human brain \citep{zhao_llm_narratives_2023, abbasiantaeb_llm_human_conversation_2024, fayyaz_llm_human_alignment_2024}. Despite this division, it has become increasingly clear that limiting these models to a few specialized areas greatly restricts their potential to fully grasp and portray the complexity of the world.
Therefore the integration of sensory modalities through technology, particularly using AI, has emerged as a compelling frontier in computer science and cognitive research \citep{murari_maluma_ml_2020, turato_yellow_door_2023}. As multimodal AI models advance, they increasingly offer innovative solutions for bridging human experiences and machine understanding across diverse sensory domains. These models, which merge information from different modalities enable machines to interpret complex real-world scenarios and provide more nuanced outputs. While recent research has predominantly focused on the intersection of audio and visual modalities, the potential for integrating taste and sound remains relatively unexplored.
Nonetheless, advances in neuroscientific and psychological research have established clear links between taste perceptions and other sensory inputs, especially auditory stimuli \citep{spence_crossmodal_2011, spence_sonic_2021, guedes_crossmodal_2023}. These investigations indicate that certain auditory characteristics can impact how tastes are perceived. For example, sounds with a low pitch are often linked with bitterness, whereas high-pitched sounds tend to be associated with sweetness \citep{crisinel_as_2010}. This rich area of crossmodal associations paves the way for cutting-edge AI applications that craft immersive sensory experiences by integrating taste and sound.
Recent progress in generative AI, notably large language models (LLMs), has showcased exceptional ability to produce coherent and contextually suitable outputs across various modalities. In music generation, models such as MusicGEN \citep{copet_simple_2024}, MusicLM \citep{agostinelli_musiclm_2023}, among others, have been designed to craft intricate musical compositions from textual cues. These models are trained extensively on datasets with a wide range of music features, enabling them to generate music that matches particular textual instructions. Nonetheless, incorporating cross-modal information into these models remains largely unexplored, offering both challenges and opportunities for future innovation.
To address these challenges, this article proposes a novel approach to incorporating taste information into music generation by refining the data provided to AI models. Building on expert studies and already existing datasets \citep{guedes_taste_2023}, a dataset has been generated that emphasizes the neuroscientific and experimental psychology knowledge underlying the relationship between taste and music. Subsequently, a generative model for music (MusicGEN) was selected for fine-tuning to assess whether the enriched data contributes effectively to the model’s internal representation. Through an online survey to evaluate the model's outputs we discovered that the model trained in this manner produces music that more accurately and coherently represents the input taste descriptions, compared to a non-fine-tuned model. To enhance visual comprehension of the research process, Figure \ref{fig:pipeline} illustrates the experimental pipeline.

The research questions guiding this study are as follows:

\begin{enumerate}
    \item Does the model fine-tuned with a neuroscientifically validated dataset produce outputs that align more coherently with the synesthetic taste effect?
    \item Can the fine-tuned model induce gustatory responses?
    \item Which underlying connections make the synesthetic effect possible?
    \item How much do emotions mediate cross-modal evaluations of the music?
\end{enumerate}

The article is organized as follows: section \ref{sec:sota} provides an overview of the background and related work in both cognitive neuroscience and computer science domains, section \ref{sec:musicgen} discusses the fine-tuned model and the datasets used in this article, section \ref{sec:m-and-m} introduces the experiment we organized to evaluate the model, section \ref{sec:results} presents the analysis of the experiment's results, section \ref{sec:discussion} discusses the results and compares them with previous literature, and section \ref{sec:conclusions} concludes with considerations on the implications of our findings and directions for future research.

\section{Background and related work}\label{sec:sota}
\subsection{Cross-modal correspondences between sounds and tastes}

The human brain demonstrates a remarkable capacity to establish consistent associations across sensory modalities, a phenomenon broadly termed crossmodal correspondences. These correspondences, systematically defined by \cite{spence_crossmodal_2011}, refer to reproducible mappings between perceptual dimensions across different sensory systems. Such associations may occur between both directly perceived and imagined stimulus attributes and can arise from shared redundancies or distinct perceptual features \citep{spence_crossmodal_2011}. One of the earliest documented examples of this phenomenon dates back to \cite{kohler_gestalt_1929}'s seminal work, where he observed that individuals tended to associate the pseudoword “baluba” with rounded shapes and “takete” with angular ones. Later research has revealed a diverse range of crossmodal correspondences encompassing nearly all combinations of sensory modalities \citep{spence_crossmodal_2011}.
While much of the foundational research emphasized pairings between visual and other sensory modalities, increasing attention has been directed towards associations involving auditory cues and the chemical senses, such as taste and smell. Auditory inputs, including environmental sounds and those emanating from food (e.g., the crunch of chips), significantly influence flavor perception and eating behavior. For example, modifying food sounds has been shown to enhance perceptions of freshness and crispness \citep{dematte_sound_apple_2014, zampini_sensory_studies_2004}, and environmental music or soundscapes can modulate meal duration, eating speed, and consumption patterns \citep[e.g.][]{mathiesen_jappet_2022}. Metaphors such as describing a melody as “sweet” or a voice as “bitter” reflect intuitive connections between auditory and gustatory modalities that have long permeated human language. For instance, the Italian musical term \textit{dolce} denotes both “sweetness” and a gentle, soft playing style \citep{knoferle_crossmodal_2012, mesz_composition_2012}. While taste-related descriptors are infrequent in musical contexts, they do appear as expressive markers on occasion. One notable example is the term \textit{{\^a}pre} (bitter), which features in Debussy’s \textit{La puerta del vino} (1913), a composition distinguished by its low pitch register and moderate dissonance. Despite these intriguing connections, systematic empirical efforts to investigate such crossmodal associations has emerged relatively recently.
\cite{holt-hansen_pms_1968, holt-hansen_pms_1976} pioneered this line of investigation by demonstrating that participants could associate the flavors of various beers with specific pitches of pure tones. For example, higher pitches (640–670 Hz) were linked to Carlsberg’s Elephant beer, whereas lower pitches (510–520 Hz) were matched to standard Carlsberg beer. Moreover, participants reported richer sensory experiences when they perceived the pitch and taste as harmonious. While replications of Holt-Hansen’s findings \citep[e.g.][]{rudmin_pms_1983} yielded mixed results—likely due to methodological limitations such as small sample sizes—they provided the groundwork for future research. 
\cite{crisinel_implicit_2009, crisinel_as_2010, crisinel_sweet_2010} expanded on these early studies using implicit association tasks to explore pitch-taste correspondences. Their findings revealed robust associations between higher-pitched sounds and sweet or sour tastes, while bitter tastes corresponded to lower-pitched sounds. Follow-up experiments using actual tastants (rather than imagined flavors) confirmed these patterns and additionally identified associations between salty tastes and medium-pitched sounds.

The researchers also examined the role of psychoacoustic properties such as timbre—characterized by spectral centroid and attack time—in shaping these associations. For example, sweet tastes were linked to piano sounds (perceived as pleasant), while bitter and sour tastes were associated with trombone timbres (perceived as unpleasant) \citep{crisinel_as_2010}. Further investigations have consistently observed associations between sweetness (and sometimes sourness) with higher-pitched sounds and bitterness with lower-pitched sounds \citep{knoferle_pm_2015, qi_perception_2020, wang_turn_2016, watson_multisensory_2017}. For instance, \cite{knoferle_pm_2015} demonstrated that both simple chord progressions and complex soundtracks were encoded with “sweet” (high-pitched) or “bitter” (low-pitched) conceptual associations. Similarly, \cite{wang_turn_2016} used a series of water-based taste solutions and MIDI-generated tones to reveal a gradient, with sour solutions paired with the highest pitches, followed by sweet, and finally bitter solutions paired with the lowest pitches. \cite{spence_crossmodal_2011} proposed three potential mechanisms underlying crossmodal correspondences: structural, statistical, and semantic. Structural correspondences derive from shared neural encoding mechanisms across sensory modalities. Statistical correspondences are shaped by regularities in the environment, such as the physical relationship between pitch and size. Semantic correspondences arise from shared descriptive language, such as the metaphorical use of terms like “sweet” across both taste and music \citep{mesz_taste_2011}. Additionally, emotional responses to stimuli can influence crossmodal correspondences. For instance, emotionally evocative stimuli, such as music, often elicit consistent crossmodal mappings \citep{mesz_frontiers_2023}. Music-color and music-painting associations are frequently predictable based on the emotional valence of the stimuli \citep{spence_multisensory_2020}. Furthermore, color can modulate music-induced emotional experiences, as shown by \cite{hauck_multisensory_2022}, who demonstrated that emotional responses to musical pieces shifted in alignment with colored lighting. Similarly, \cite{galmarini_food_2021} found that the emotional tone of background music could shape the sensory experience of drinking coffee. The emotional responses evoked by music and taste could serve as a link for crossmodal associations by aligning the emotional qualities of both stimuli. The emotional valence of both the music and the taste may share similar underlying affective dimensions, such as pleasantness or unpleasantness, which could drive the association. Music and taste can elicit emotional reactions, and when these emotional responses are congruent, it is likely that the brain establishes connections between them, leading to a crossmodal association based on shared emotional experiences. In conclusion, crossmodal correspondences offer a compelling framework for investigating the interconnected nature of sensory perception. Moreover, these findings highlight the potential for using auditory stimuli to influence gustatory perception. For example, restaurants might design soundscapes to enhance specific taste qualities or improve the overall dining experience. 

\subsection{Cross-modal generative models}

In recent years, cross-modal generative models have advanced significantly, inspired by an increasing interest in developing systems capable of seamlessly integrating and translating information across diverse sensory modalities. This evolution is driven by the increasing capabilities of artificial intelligence, particularly within the realm of generative models, which have demonstrated potential in producing coherent and contextually relevant outputs across a multitude of domains. 
The advancement of cross-modal generative models is grounded in foundational research within the disciplines of cognitive neuroscience and experimental psychology, which have long investigated the interactions among different sensory modalities. These models endeavor to emulate the human faculty of perceiving and interpreting multisensory information, a process that is inherently complex and nuanced. By utilizing large-scale datasets and advanced machine learning techniques, researchers have initiated the creation of models capable of generating outputs that reflect the intricate interrelations among modalities such as vision, sound, and taste.
Several notable multimodal generative models have emerged, illustrating the substantial capabilities inherent within this domain. Text-to-image generation models, such as DALL·E \citep{ramesh_zero-shot_2021} and Stable Diffusion \citep{rombach_diffusion_2022}, are capable of rendering detailed images from textual descriptions. Text-to-audio models, including MusicLM \citep{agostinelli_musiclm_2023}, translate text prompts into music or soundscapes, presenting intriguing possibilities for the fields of entertainment and virtual environments. Although still at a nascent stage, text-to-video generation (generating both video and audio) is anticipated to offer significant benefits for media content production and simulation environments \citep{singer_make--video_2022}.
In contrast, image-to-text models \citep{radford_learning_2021, li_blip_2022, alayrac_flamingo_2022} transform visual data into descriptive narratives, thereby facilitating tasks such as automated captioning and providing assistance to individuals with visual impairments. Audio-to-text models, which have been widely implemented in speech-to-text applications, have historically served the domains of transcription and virtual assistance \citep{bahar_ieee_2019}. Recent developments in generative models have enabled more nuanced and context-sensitive analyses of spoken language.

An emerging but relatively underexplored field in multimodal AI is emotional awareness integration. Although significant work has gone into identifying emotions within just one modality \citep{poria_review_2017}, there is growing interest in synthesizing data from multiple modalities \citep{poria_review_2017, zhao_affective_2019}. This multimodal strategy is beneficial because integrating data from various sources enhances emotion recognition capabilities and opens up to new possibilities which are not possible at the moment with just a text-based approach as in \cite{boscher_sense-lm_2024}. However, research into how different modalities correlate based on emotions applied to computer science has been rather limited. Recent developments, such as those discussed in \citep{zhao_emotion-based_2020}, demonstrate viable ways of linking visual and auditory data through an emotional valence-arousal latent space using supervised contrastive learning methods. This advancement enables a more detailed and flexible representation of emotional states than the traditional concept of distinct emotions, capturing the intricate and nuanced nature of human emotions and offering a broader comprehension of their interactions across diverse sensory stimuli.
This approach aligns with the broader goal of creating AI systems that are not only technically proficient but also capable of understanding and responding to human emotions in a meaningful way.
Despite these advancements, several challenges remain in the development of cross-modal generative models. One significant hurdle is the need for comprehensive datasets that encompass the full spectrum of sensory experiences. Current datasets often lack diversity, limiting the ability of models to generalize across different contexts and populations. Additionally, the complexity of human emotions and their influence on sensory perception presents a formidable challenge, requiring sophisticated models that can accurately capture and interpret these nuances.
The future of cross-modal generative models involves ongoing improvements and enhancements, particularly in terms of developing their emotional intelligence and broadening their range of applications. By tackling present constraints and seizing the possibilities unlocked by multimodal integration, researchers can advance towards AI systems that deliver more engaging and tailored experiences, effectively closing the divide between human perception and machine-generated results.

\section{MusicGEN}\label{sec:musicgen}

In this study, MusicGEN -- a cutting-edge generative model specifically engineered for music -- was fine-tuned and then used to generate music compositions. The fine-tuning process was pivotal in adapting the model to our research context, which centers on exploring the nuanced interplay between musical compositions and sensory-gustatory responses. To facilitate this adaptation, we utilized a patched version of the Taste \& Affect Music Database \citep{guedes_taste_2023}. This database originally encompassed a diverse range of musical pieces, each accompanied by evaluations reflecting gustatory and emotional responses. We enhanced this foundational dataset by incorporating descriptive captions for each audio file, meticulously crafted by the authors to include detailed information on the intended flavors and emotional qualities associated with each musical piece. In addition, these captions encompassed relevant audio metadata such as tempo, key, and instrumentation. This enhancement was designed to provide richer contextual information to the model, with the aim of generating music that more accurately mirrors the complexities inherent in taste descriptions and emotional nuances.
In our exploration of multimodal generative models for music synthesis, we critically evaluated several candidates, including MusicLM, Riffusion \citep{forsgren_riffusion_2022}, and MusicGEN. MusicLM, developed by Google, presents a robust architecture for generating music from textual prompts; however, its closed-source nature imposes significant restrictions on customization and adaptability, rendering it less suitable for our specific research objectives Riffusion, while innovative in its approach to music generation through the utilization of Stable Diffusion, was excluded from consideration due to inherent limitations such as the necessity of converting audio into spectrograms that introduces additional computational overhead and its inability to maintain coherent long-term audio sequences, as discussed in \cite{huang_make--audio_2023-1}. Unlike Riffusion, MusicGEN's Transformer-based architecture supports the retention of internal states, enabling the model to produce more coherent and contextually relevant   MusicGEN, developed by Meta, is an open-source model that permits extensive modifications and fine-tuning, making it a far more appropriate choice for our study's aims. 
 musical outputs. Thus, MusicGEN was selected for its optimal balance of accessibility, flexibility, and capacity to generate coherent music that is in line with taste descriptors.
MusicGEN is characterized as a state-of-the-art autoregressive transformer-based model \citep{vaswani_attention_2017}, specifically designed to generate high-quality music at a sampling rate of 32 kHz. The model operates by conditioning on either textual or melodic representations, which empowers it to produce coherent musical pieces that are in harmony with the provided input context. Its architecture employs a single-stage language model that leverages an efficient codebook interleaving strategy, facilitating the simultaneous processing of multiple discrete audio streams. This innovative approach is made possible through the integration of an EnCodec audio tokenizer \citep{defossez_high_2022}, which quantizes audio signals into discrete tokens, thus enabling high-fidelity reconstruction from a low frame rate representation. The design of the model incorporates Residual Vector Quantization (RVQ) \citep{zeghidour_soundstream_2021}, resulting in several parallel streams of discrete tokens derived from distinct learned codebooks.

The capability of MusicGEN to generate music is further enhanced by its proficiency in performing both text- and melody-conditioned generation. This dual conditioning mechanism allows the model to maintain fidelity to the textual descriptions while ensuring that the generated audio remains coherent with the specified melodic structure. However, it is important to acknowledge that, despite its numerous strengths, the model does encounter limitations regarding fine-grained control over the adherence of the generated output to the conditioning inputs.
To adapt MusicGEN for our specific task of generating music based on taste descriptors, we undertook a comprehensive fine-tuning process.
In our fine-tuning endeavors, we opted to utilize the smaller variant of MusicGEN, comprising 300 million parameters, to ensure efficient training while still maintaining sufficient representational capacity. The fine-tuning process was conducted over 30 epochs, employing a batch size of 16 and a learning rate set at $1.0 \times 10^{-4}$ adjusted accorting to a cosine schedule. The AdamW optimizer was used, featuring a weight decay of 0.01, and the training process involved 2000 updates per epoch. This specific configuration was carefully chosen to strike a balance between convergence speed and overall model performance.
The fine-tuning was executed on the "Blade" cluster at the Department of Information Engineering (DEI) at the University of Padua, utilizing two NVIDIA RTX3090 GPUs, each equipped with 24 GB of VRAM.

\subsection{Dataset}

MusicGEN has been originally trained on a non-public dataset of 20k hours of music collected by Meta. This kind of dataset is particularly effective to make the model figure out, after a training period, the underlying structures embedded in musical artifacts, on the other side the music generated by the model could lack in specificity or could have some kind of bias. While this study does not focus primarily on biases, of generating functional music requires creating compositions that adhere to specific attributes.This is where fine-tuning comes into play::it allows us to refine the model by focusing on a specific dataset where particularconditions are met. To fine-tune the model so that it is aware of the correlations between auditory and gustatory experiences, we created a patched version of the taste \& affect music database by \cite{guedes_taste_2023}.

The Taste \& Affect Music Database \citep{guedes_taste_2023} was born as a resource for investigating the intricate relationships between auditory stimuli and gustatory perceptions. This dataset comprises 100 instrumental music tracks, meticulously curated to encapsulate a diverse range of emotional and taste-related attributes. Each musical piece within the database is accompanied by subjective rating norms that reflect participants' evaluations across various dimensions, including basic taste correspondences, emotional responses, familiarity, valence, and arousal.
The selection of musical stimuli was guided by the objective of establishing clear associations between auditory and gustatory attributes. The tracks were chosen to represent fundamental taste categories—sweetness, bitterness, saltiness, and sourness—allowing researchers to explore how these tastes can be conveyed through music.
Each participant provided ratings on the music tracks using a series of self-report measures that assessed mood, taste preferences and musical sophistication. This multi-dimensional approach to data collection facilitated a nuanced understanding of how individual differences in taste perception and emotional responses can influence the evaluation of musical stimuli.
To adapt the dataset for fine-tuning, we generated captions for each music sample. These captions specify musical elements such as tempo, key and instrumentation. Furthermore, we incorporated keywords extracted from the original Taste \& Affect Music Database, designating each sample as representative of one or more taste categories only if its score exceeded 25\% in the original dataset.

\subsection{Generated Dataset}\label{sec-gen-data}

We then tested the model prompting it to infer different kind of music, at first few qualitative attempts were made to assess the correspondence between the prompted text and the model's output. In particular we performed a qualitative stress test varying musical genre asking, with many different prompts, for classical, ambient and jazz music. We found that the generated audio matched with varying quality the prompt with the exception of classical music, where the models (both the base and the fine-tuned version) tend to disattend the prompt with non-classical music, one reason could be the fact that the 20k hours training dataset of the MusicGEN model comprehends just a small percentage of classical music, while the corpora better represents other genres such as jazz and ambient. The ambient genre showed to be the most neutral one and adapt to generate music suited to be evaluated by subject without being conditioned by the genre, hence we kept specifing this genre in the successive prompts to avoid other genre biases during the output evaluation.
Following a qualitative assessment, we created a dataset using both the original and fine-tuned models. Four prompts were developed, corresponding to each taste under study, with the structured format: $\langle$TASTE$\rangle$ \textit{music, ambient for fine restaurant}, where $\langle$TASTE$\rangle$ represents sweet, bitter, sour, and salty. Each model produced a total of 100 pieces, each lasting 15 seconds. Of these, 25 were generated using the salty prompt, 25 using the sweet prompt, 25 using the bitter prompt, and 25 using the sour prompt. To compare outputs, we adopted standard metrics to evaluate the fine-tuned model in relation to the base version, specifically measuring the Fr\'echet Audio Distance (FAD) \citep{kilgour_interspeech_2019} between the training dataset and the outputs of both models when given the same prompt.
The evaluation has been performed adopting the fadtk implementation \citep{gui_icassp_2024} using VGGish embeddings as in the original MusicGEN paper \citep{diwakar_springer_2024}, in addition with the EnCodec ones, since the model is based on such encoder we think that this metric should better match the internal representation of the model.

\begin{table}[h]
    \centering
    \begin{tabular}{lcc}
        \toprule
        Model & VGGish & EnCodec \\
        \midrule
        Base       & $3.184$ & $121.513$ \\
        Fine-Tuned & $2.579$ & $107.594$ \\
        \bottomrule
    \end{tabular}
    \caption{FAD evaluation results using VGGish and EnCodec embeddings.}
    \label{tab:fad_results}
\end{table}

The evaluation results shown in Table \ref{tab:fad_results} display that the music generated by fine-tuned model better matches with the reference dataset, despite could be an expected result that a fine-tuned model generates music more similar to the training dataset than its non fine-tuned version it is important to denote that the training dataset was just 1 hour length and very specific.

\section{Material and Methods}\label{sec:m-and-m}
The subjective evaluation of the fine-tuned model was conducted through an online survey administered through PsyToolkit, a widely used platform for psychological research \citep{stoet_psytoolkit_2017}. The survey was structured to gather participants' opinions on the gustatory effects induced by the fine-tuned model compared to the non-fine-tuned version. Participants were recruited through various online channels to ensure a diverse demographic representation. 

The listening tasks consisted of two distinct types. In the first task, participants were asked to express their preference between two audio files generated by the two models. In the second task, they quantified their perceptions and emotional responses to each piece of music. Specifically, participants rated the flavors they perceived using a graduated scale from 1 to 5 for four primary taste categories: salty, sweet, bitter, and sour. Additionally, they rated their emotional responses on various non-gustatory parameters, including happiness, sadness, anger, disgust, fear, surprise, hot and cold, using the same graduated scale. This survey design allowed the collection of both quantitative and qualitative data, facilitating a comprehensive analysis of the relationship between music and sensory experiences.

All materials, including the patched database, survey instruments, and detailed instructions for the fine-tuning process, are available for reproducibility and further research.

\subsection{Participant Selection and Demographic Data}

Participants were recruited through a combination of online platforms and local community outreach, ensuring a diverse sample reflective of the general population. A total of 111 individuals participated in the study, comprising 61 males, 46 females, 2 individuals identified as other, and 2 who did not specify their gender. The mean age of the participants was 32 years (with a minimum age of 19 and a maximum age of 75). Along with gender and age, we collected a self-evaluation of both the auditory (38 professionals, 43 amateurs, 30 not-experienced) and the gustatory experience (1 professional, 44 amaterus, 66 not-experienced), the ethnicity and the type of audio device used to participate in the survey (headphones, speakers or HiFi stereo).

This study was conducted in accordance with the ethical principles outlined in the Declaration of Helsinki (most recently amended in 2024), ensuring respect for participants' rights, safety, and well-being. Prior to participation, all individuals provided an informed consent after receiving a detailed explanation of the study's objectives, procedures, and voluntary nature. Given the non-invasive nature of the survey, the study was classified as zero-risk research according to the ethical self-assessment guidelines of the Committee for Ethical Research (CER) of the University of Trento, and thus did not require additional ethical approval.

\subsection{Experiment Design}

The online survey was structured to explore the relationship between auditory stimuli and taste perception. Initially, participants selected their preferred language to complete the survey, with options available in both English and Italian. 
Following the language selection, participants engaged in a series of listening tasks.

\subsubsection{Task A}
\begin{figure}
     \centering
     \begin{subfigure}[b]{0.45\textwidth}
         \centering
         \includegraphics[width=\linewidth]{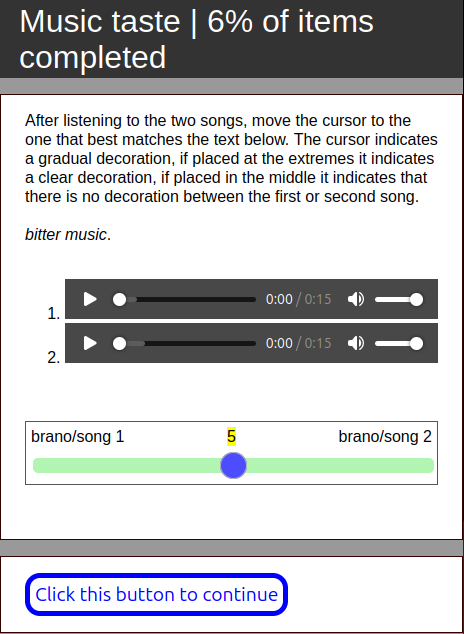}
         \caption{Survey's task A interface, where participants express song preference.}
         \label{fig:task1}
     \end{subfigure}
     \hfill
     \begin{subfigure}[b]{0.45\textwidth}
         \centering
         \includegraphics[width=\linewidth]{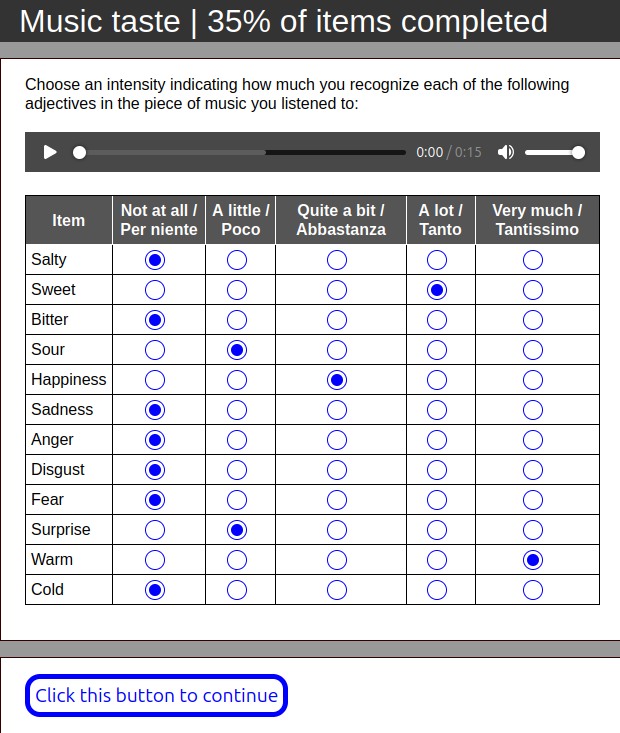}
         \caption{Survey's task B interface.}
         \label{fig:task2}
     \end{subfigure}
        \caption{User interfaces for the two survey tasks. Panel (a) presents task A, where participants rate their preference between two generated songs, while panel (b) shows task B, which evaluates the perception of different musical characteristics.}
        \label{fig:survey-tasks}
\end{figure}

The first task involved the presentation of two audio clips, each associated with a specific taste category (sweet, salty, bitter, and sour), see Figure \ref{fig:task1}. Participants were made aware of the taste category through a text indicating if the music was supposed to be perceived as sweet, salty, bitter or sour, then they had to listen attentively to both clips before indicating which of the stimuli was the most coherent with the given text by moving a cursor along a scale ranging from 0 to 10. This scale allowed for a nuanced expression of preference, where a position of 0 indicated a strong preference for the first audio clip, a position of 10 indicated a strong preference for the second, and a position of 5 signified no preference between the two.
To mitigate potential biases, the order of taste categories and audio clips was randomized for each participant. Each audio clip was generated by either the fine-tuned model or the base model as specified in section \ref{sec-gen-data}, although the participants were not informed of the specific model used for each clip. This design choice aimed to improve the robustness of the findings by controlling for model-related effects.
Participants completed a total of five listening tasks, each featuring different audio clips corresponding to randomly assigned taste categories.

\subsubsection{Task B}
Following the five items of task A, participans where presented with three more items, each one including one single audio stimulus and an evaluation based on the list of 12 adjectives-words, see Figure \ref{fig:task2}, this list includes the six basic emotions by \cite{ekman_cognition_1992}, the four basic tastes and temperature feeling (hot, cold), for each of these words participants used a scale from 1 to 5 to quantify their perception (where 1 means not at all and 5 means a lot). We considered these adjective-words to study eventual correlations between tastes and other domains such as emotions and thermal perception.
This evaluation allowed participants to articulate the extent to which they recognized each adjective in relation to the music they had just listened to.

\section{Results}\label{sec:results}
In this section have been reported the most meaningfull results by a deeper analysis that we conducted. The full analysis along with the scripts used to generate the results can be found at the following website: \url{https://matteospanio.github.io/multimodal-symphony-survey-analysis/}.

\subsection{Task A analysis}
The objective of analyzing these results is to determine whether one model is consistently judged as more accurate than the other in generating music associated with the given prompt. Therefore, we evaluated whether the scores systematically favor one model over the other.

At first, due to the random order of the stimuli presentation, we normalized the scores attributed in task A ordering the preference according to the score function $S$ as defined in Equation \ref{eq:score-fn}. This procedure allows us to interpret scores from 0 to 4 as a preference for the base model, scores between 6 and 10 as a preference for the fine-tuned model, and scores of 5 are treated as neutral.

\begin{equation}\label{eq:score-fn}
    S(x, m) =
    \begin{cases} 
        x, & \text{if } m = \text{right} \\
        10 - x, & \text{if } m = \text{left}
    \end{cases}
\end{equation}
where $x \in \{ n \in \mathbb{N} \mid n \leqslant 10 \}$, and \( m \) can take the values ``right'' or ``left'', according to the position on the survey form of the stimulus generated with the fine-tuned model.

\begin{figure}[h!]
\begin{center}
\includegraphics[width=10cm]{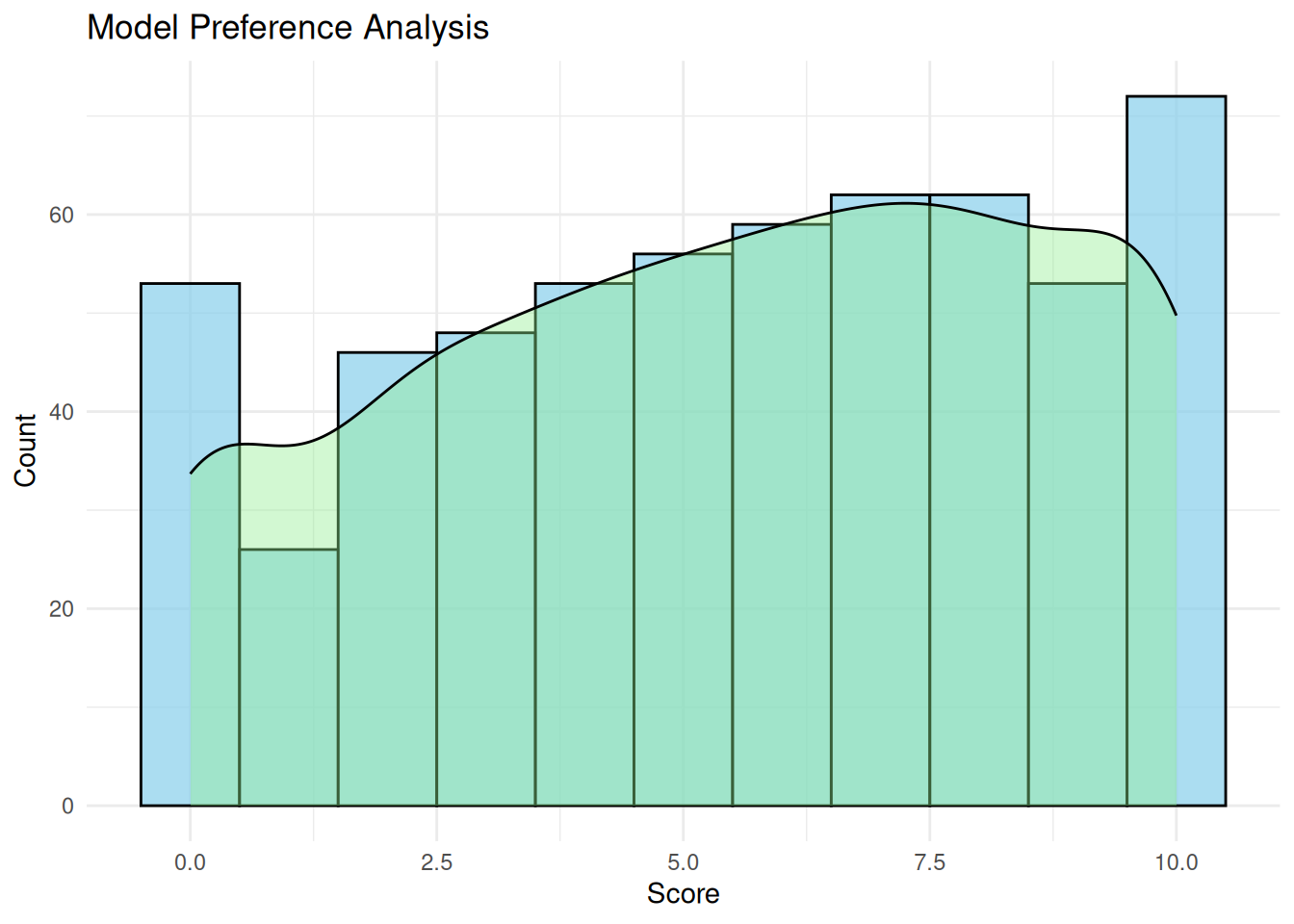}
\end{center}
\caption{Distribution of the task A results, where a score of zero means a strong preference for the base model, while a score of 10 means a strong preference for the fine-tuned model.}\label{fig:model-preference}
\end{figure}

An histogram of the participants' ratings is shown in Figure \ref{fig:model-preference}, where a preference for the fine-tuned model is evident, due to the right-skewed statistical distribution.
After a Shapiro-Wilk test with $p < 0.05$ that determined the non normal distrubution of the data, we opted for a Wilcoxon signed-rank test which gave a statistically significant result supporting the hypothesis that the median score is greater than 5 ($p<0.001, W=73966$).
Furthermore we continued with a post-hoc analysis performing a Wilcoxon test for each taste group (sweet, sour, bitter, salty) applying a Bonferroni correction to adjust for the multiple comparisons and control the family-wise error rate. 

\begin{table}[ht]
    \centering
    \begin{tabular}{lccc}
        \toprule
        \textbf{taste} & \textbf{$p$}  & \textbf{W} & \textbf{adjusted $p$} \\
        \midrule
        bitter & $0.003$  & $5573$ & $0.013$\\
        salty  & $0.996$  & $2784$ & $1.000$\\
        sour   & $0.007$  & $4770$ & $0.030$\\
        sweet  & $<0.001$  & $5320$ & $<0.001$\\
        \bottomrule
    \end{tabular}
    \caption{$p$-values and adjusted $p$-values resulting from the Wilcoxon test for different taste attributes.}
    \label{tab:multi-wilcoxon}
\end{table}

As can be seen from the results of the test reported in Table \ref{tab:multi-wilcoxon} all audio samples generated by the fine-tuned model but \textit{salty}, are statistically chosen as better than the base model. We then performed the opposite hypothesis test to test just the mean of the salty group of samples, the result confirm a median score lower than 5, meaning that the base model is overall preferred in the case of salty text suggestions ($p\approx 0.003, W=2784$).

\subsection{Task B analysis}
To investigate whether different prompts and adjectives resulted in significantly different ratings assigned by participants, and to examine the interaction between taste, emotions, and thermal perception, we first conducted an Analysis of Variance (ANOVA), followed by a factor analysis. The ANOVA model is defined as follows:

\begin{equation}\label{eq:anova}
\text{value} \sim \text{prompt} \times \text{adjective} + \text{hearing\_experience} + \text{eating\_experience} + \text{gender}
\end{equation}
where $\times$ denotes an interaction effect between factors, \textit{value} represents the score assigned by the participant to a specific \textit{adjective}, \textit{prompt} refers to the designated taste category used during stimulus generation, \textit{gender} corresponds to the participant's self-reported gender, while \textit{hearing\_experience} and \textit{eating\_experience} indicate the participant's self-assessed expertise in auditory and gustatory tasks, respectively.

The dataset was filtered to include only participants identified as Male or Female, excluding other genres and excluding also participants classified as Professional Eaters due to insufficient representation of these categories.

\begin{table}[ht]
    \centering
    \begin{tabular}{lccccc}
        \toprule
        \textbf{Factor} & \textbf{Df} & \textbf{Sum Sq} & \textbf{Mean Sq} & \textbf{F value} & \textbf{Pr($>$F)} \\
        \midrule
        prompt               & $3$    & $29.402$   & $9.801$   & $8.739$   & $<0.001$ \\
        adjective            & $11$   & $188.478$  & $17.134$  & $15.279$  & $<0.001$ \\
        hearing experience   & $2$    & $37.299$   & $18.650$  & $16.630$  & $<0.001$ \\
        eating experience    & $1$    & $0.711$    & $0.711$   & $0.634$   & $0.426$ \\
        gender               & $1$    & $0.069$    & $0.069$   & $0.061$   & $0.804$ \\
        prompt:adjective     & $33$   & $214.757$  & $6.508$   & $5.803$   & $<0.001$ \\
        \bottomrule
    \end{tabular}
    \caption{Results of the ANOVA test.}
    \label{tab:anova_results}
\end{table}
The ANOVA results (see Table \ref{tab:anova_results}) show a significant effect of both prompt and adjective, with an even stronger effect for their interaction. In other words, the prompt influences participants' ratings across the different adjectives-words in the semantic scale.
Also hearing experience shows to be relevant in order to evaluate the audio stimuli, whereas neither eating experience nor participant's gender influenced the stimuli evaluations of Task B.
A post-hoc analysis was then conducted on the significative factors by means of the Tukey's Honest Significant Difference (HSD) test.
Tables \ref{tab:tukey_prompts} and \ref{tab:tukey_adjectives} list the combinations of, respectively, prompts and adjectives that show statistically significant differences. Notably the sour prompt received higher evaluations compared to other ones. Table \ref{tab:tukey_adjectives} instead highlights that \textit{anger} and \textit{disgust} received lower values overall, while \textit{hot, cold} and \textit{sad} received the highest evaluations.

\begin{table}[ht]
    \centering
    \begin{tabular}{lcccc}
        \toprule
        \textbf{Comparison} & \textbf{diff} & \textbf{lwr} & \textbf{upr} & \textbf{p adj} \\
        \midrule
        sour - bitter & $0.153$ & $0.027$ & $0.280$ & $0.009$ \\
        sour - salty  & $0.194$ & $0.066$ & $0.322$ & $<0.001$ \\
        sweet - sour  & $-0.193$ & $-0.321$ & $-0.065$ & $<0.001$ \\
        \bottomrule
    \end{tabular}
    \caption{Tukey test results for different prompts with a $p$-value lower than 0.05.}
    \label{tab:tukey_prompts}
\end{table}

\begin{table}[ht]
    \centering
    \begin{tabular}{lcccc}
        \toprule
        \textbf{Comparison} & \textbf{diff} & \textbf{lwr} & \textbf{upr} & \textbf{p adj} \\
        \midrule
        bitter - anger   & $0.420$  & $0.143$  & $0.696$  & $<0.001$ \\
        cold - anger     & $0.477$  & $0.201$  & $0.754$  & $<0.001$ \\
        hot - anger      & $0.531$  & $0.255$  & $0.808$  & $<0.001$ \\
        sad - anger      & $0.576$  & $0.300$  & $0.853$  & $<0.001$ \\
        sweet - anger    & $0.372$  & $0.095$  & $0.648$  & $<0.001$ \\
        disgust - bitter & $-0.633$ & $-0.910$ & $-0.357$ & $<0.001$ \\
        happy - bitter   & $-0.309$ & $-0.585$ & $-0.032$ & $0.013$ \\
        surprise - bitter & $-0.285$ & $-0.561$ & $-0.008$ & $0.036$ \\
        disgust - cold   & $-0.690$ & $-0.967$ & $-0.414$ & $<0.001$ \\
        happy - cold     & $-0.366$ & $-0.642$ & $-0.089$ & $<0.001$ \\
        sour - cold      & $-0.285$ & $-0.561$ & $-0.008$ & $0.036$ \\
        surprise - cold  & $-0.342$ & $-0.618$ & $-0.065$ & $0.003$ \\
        fear - disgust   & $0.432$  & $0.155$  & $0.708$  & $<0.001$ \\
        happy - disgust  & $0.324$  & $0.047$  & $0.600$  & $0.007$ \\
        hot - disgust    & $0.744$  & $0.468$  & $1.021$  & $<0.001$ \\
        sad - disgust    & $0.789$  & $0.513$  & $1.066$  & $<0.001$ \\
        salty - disgust  & $0.468$  & $0.192$  & $0.745$  & $<0.001$ \\
        sour - disgust   & $0.405$  & $0.128$  & $0.681$  & $<0.001$ \\
        surprise - disgust & $0.348$  & $0.071$  & $0.624$  & $0.002$ \\
        sweet - disgust  & $0.585$  & $0.309$  & $0.862$  & $<0.001$ \\
        hot - fear       & $0.312$  & $0.035$  & $0.588$  & $0.012$ \\
        sad - fear       & $0.357$  & $0.080$  & $0.633$  & $0.001$ \\
        hot - happy      & $0.420$  & $0.143$  & $0.696$  & $<0.001$ \\
        sad - happy      & $0.465$  & $0.189$  & $0.742$  & $<0.001$ \\
        sour - hot       & $-0.339$ & $-0.615$ & $-0.062$ & $0.003$ \\
        surprise - hot   & $-0.396$ & $-0.672$ & $-0.119$ & $<0.001$ \\
        salty - sad      & $-0.321$ & $-0.597$ & $-0.044$ & $0.008$ \\
        sour - sad       & $-0.384$ & $-0.660$ & $-0.107$ & $<0.001$ \\
        surprise - sad   & $-0.441$ & $-0.717$ & $-0.164$ & $<0.001$ \\
        \bottomrule
    \end{tabular}
    \caption{Tukey test results for different adjectives with a $p$-value lower than 0.05.}
    \label{tab:tukey_adjectives}
\end{table}

\begin{figure}
     \centering
     \begin{subfigure}[b]{0.45\textwidth}
         \centering
         \includegraphics[width=\linewidth]{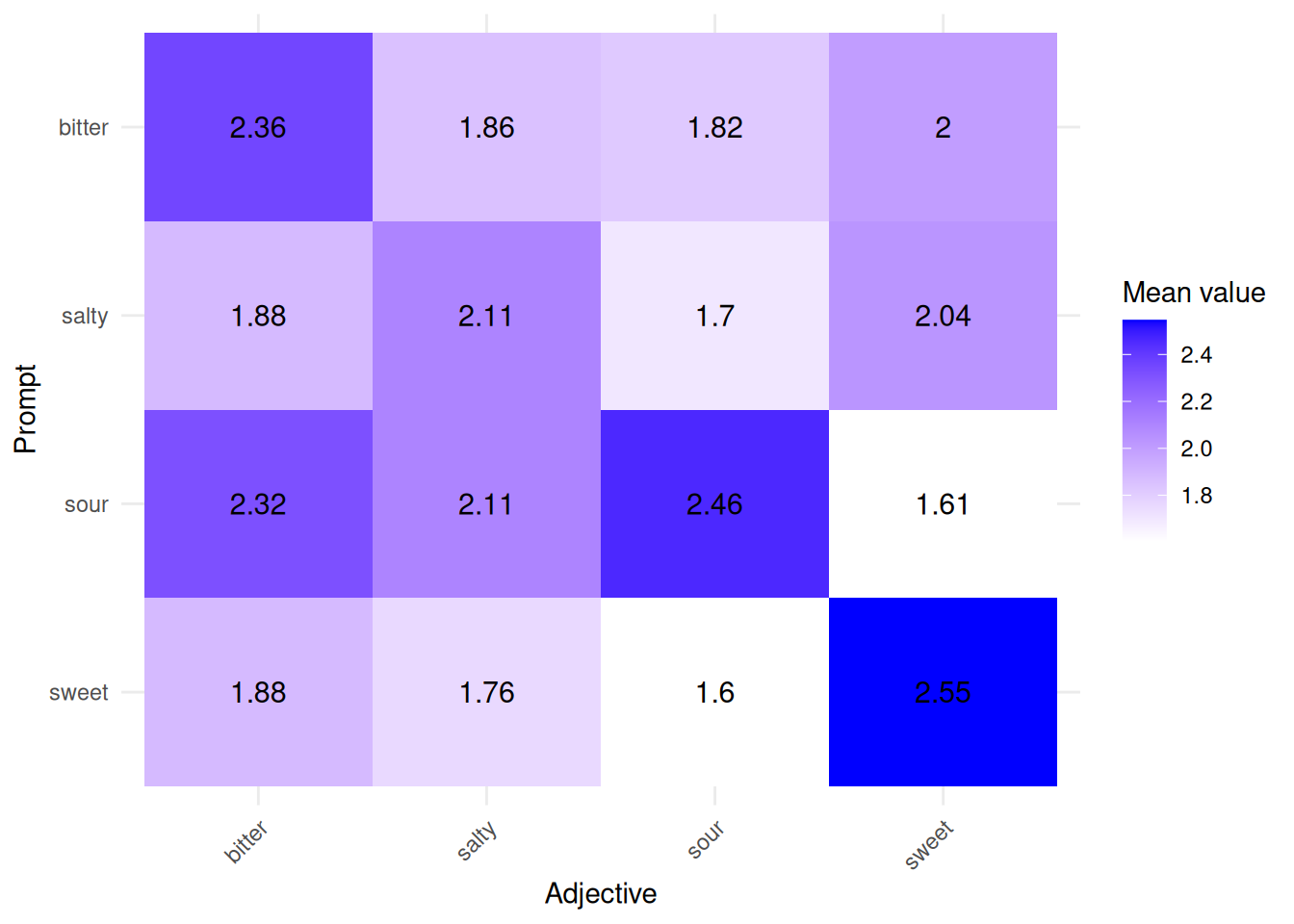}
         \caption{Heatmap of perceived taste in correspondence of the intended one.}
         \label{fig:interaction-matrix-1}
     \end{subfigure}
     \hfill
     \begin{subfigure}[b]{0.45\textwidth}
         \centering
         \includegraphics[width=\linewidth]{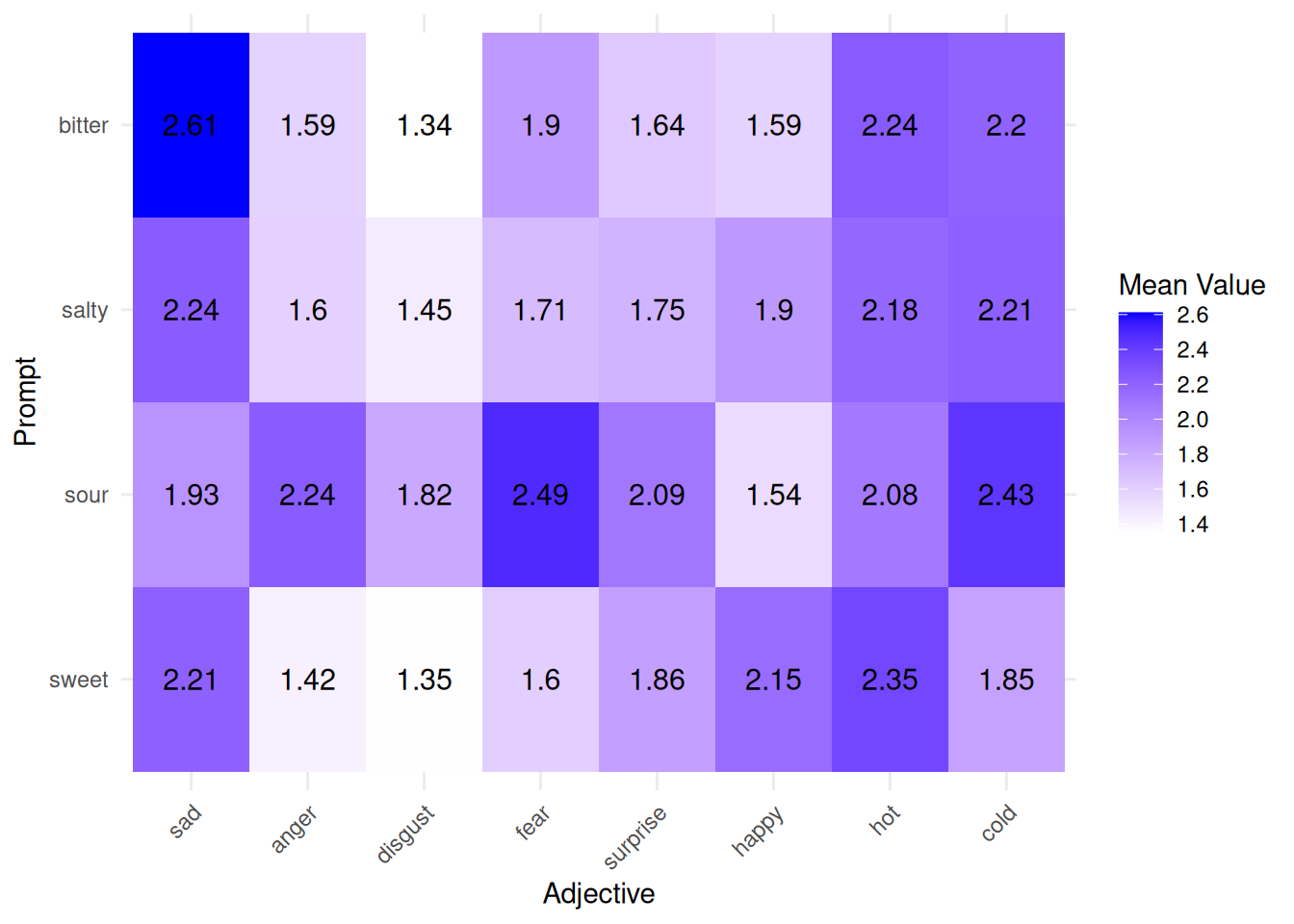}
         \caption{Heatmap of perceived emotional response in correspondence of the suggested taste.}
         \label{fig:interaction-matrix-2}
     \end{subfigure}
        \caption{Visualization of the relationship between taste prompts and their perceived characteristics. Panel (a) illustrates the alignment between intended taste prompts and the corresponding perceived taste intensities, while panel (b) presents the emotional responses elicited by each taste prompt. Color intensity represents the mean reported values.}
        \label{fig:interaction-matrices}
\end{figure}

The \textit{prompt-adjective} interaction can be seen in Figure \ref{fig:interaction-matrices}. In particular \ref{fig:interaction-matrix-1} shows the mean value assigned to each taste adjective by their prompt, we can clearly see the major diagonal emerge by the matrix, which means that the mean value assigned to the adjective that matches the prompt of each sound is the highest. The rest of the interaction between adjectives and prompts can be seen in \ref{fig:interaction-matrix-2}, a deeper analysis of emotional aspect assigned to the sounds is presented in section \ref{sec:discussion}.

The Tukey test results for the hearing experience interaction show that amateur listeners tend to give significantly higher ratings compared to professionals (\(\text{diff} = 0.23\), \( p < 0.0001\)) and the not-experienced people (\(\text{diff} = 0.17\), \( p = 0.0003\)).

To investigate the connections between sensory qualities and emotional states, we performed a factor analysis. The scree test indicated that 4 factors were optimal. Consequently, we employed a factor analysis with oblique axis rotation and the maximum likelihood method, utilizing the \textit{psych} R package by \cite{r_psych_cran}. The loadings obtained are presented in Table \ref{tab:factor-loadings}, showing the degree to which each variable contributes to the identified factors, thus offering insights into the data's underlying structure. Each of this factors is clearly characterized: the first one is about negative valence adjectives and groups together bitterness and sourness, factor two is strongly aligned with sweetness which also correlates with happiness, hotness and, a little, with sadness, factor three reaches highest scores in \textit{hot} and \textit{cold} defining a temperature dimension and factor four binds together saltiness, happiness with surprise.

\begin{table}[ht]
    \centering
    \begin{tabular}{lcccc}
        \toprule
        & \textbf{Factor 1} & \textbf{Factor 2} & \textbf{Factor 3} & \textbf{Factor 4} \\
        \midrule
        salty    &          & $-0.231$ & $ 0.111$  & $0.535$  \\
        sweet    &          & $0.992$  &          &          \\
        bitter   & $0.502$  &          &          &          \\
        sour     & $0.385$  & $-0.128$ & $ 0.178$  & $0.226$  \\
        happy    & $-0.197$ & $0.302$  & $-0.132$ & $0.492$  \\
        sad      & $0.292$  & $0.259$  & $ 0.236$  &          \\
        anger    & $0.779$  &          &          &          \\
        disgust  & $0.694$  &          &          & $-0.133$ \\
        fear     & $0.662$  &          & $ 0.133$  & $0.113$  \\
        surprise &          &          & $ 0.120$  & $0.526$  \\
        hot      & $0.140$  & $0.361$  & $-0.458$ & $0.267$  \\
        cold     &          &          & $ 0.882$  &          \\
        \midrule
        proportion variance & $0.174$ & $0.113$ & $0.096$ & $0.081$ \\
        cumulative variance & $0.174$ & $0.287$ & $0.382$ & $0.463$ \\
        \bottomrule
    \end{tabular}
    \caption{Loadings resulting from the factor analysis with 4 factors: Factor 1 includes negative valence emotions and tastes, Factor 2 is primarily associated with sweetness and other positive valence traits, Factor 3 is largely linked to temperature and shows no strong correlation with any specific taste, and Factor 4 combines saltiness and surprise with happiness.}
    \label{tab:factor-loadings}
\end{table}

\section{Discussion}\label{sec:discussion}
The findings of this study reveal that the music produced by our model refined with a dataset confirmed by psychological synesthetic research can indeed evoke synesthetic effects. Additionally, the music is not merely perceived generically as tasty; the model can be specifically prompted with particular taste attributes which, according to ANOVA tests, are often identified by listeners.

Regarding the first research question, focused on evaluating the ability of the fine-tuned model to generate audio that accurately describes the investigated flavors, the findings reveal that the fine-tuned model produced music that is more coherently aligned with the taste descriptions for sweet, sour, and bitter categories compared to the non-fine-tuned model. This indicates that the integration of gustatory information into the music generation process was effective, enhancing the model's ability to capture the sensory nuances associated with various tastes. However, music intended to represent salty flavors was less effectively captured by the fine-tuned model than by the base model.
Although the overall assessment shows that the fine-tuned model aligns better with the synesthetic effect through both objective and subjective evaluations, the salty music was better represented by the base model. One possible explanation for this phenomenon could be attributed to biases in the specified musical genre within the prompts and the dataset used for fine-tuning, where salty music is underrepresented compared to other categories. Notably, in the dataset provided by \cite{guedes_taste_2023}, the compositions are more frequently perceived as sweet, and many of those scoring well in the salty category also exhibit sweetness. Furthermore, the Fr\'echet distance based on both used embeddings suggests that the music generated by the fine-tuned model is perceptually more similar to that generated by the other model \citep{gui_icassp_2024}. This implies that the sonic characteristics of the tracks in the dataset used for fine-tuning do not adequately reflect saltiness. According to \cite{wang_multisensory_2021}, short and articulated sounds, along with steady rhythm, can evoke this sensation. The average beats per minute (BPM) of our dataset is 111 (not particularly fast), and recurring keywords include ``small emotions'' and ``ambient.'' It should be noted that ambient music is often used as background music, lacking prominent peaks in energy, timbre, and/or aggressive speed \citep{scarratt_nsr_2023}. Therefore, we conclude that while the fine-tuning was successful, the reference dataset requires further study and enrichment with music that better represents saltiness, not limited to the ambient genre.

To explore the second, third, and fourth research questions -- whether the fine-tuned model can induce gustatory responses, which underlying connections make the synesthetic effect possible, and how much emotions mediate cross-modal evaluations of music -- the study examined the extent to which the music generated by the fine-tuned model elicited synesthetic taste perceptions in participants, with a particular focus on emotional correlations. The findings indicate that the music did indeed evoke gustatory sensations, with correlations showing that positive valence emotions are associated with positive valence tastes and vice versa, while temperature also plays a significant role in these correlations. Although emotions explain a substantial portion of the correlations, the factor analysis revealed that the four factors accounted for less than 50\% of the total variance.
The ANOVA test results confirm that participants perceived taste suggestions guided by an undergoing logic rather than randomly. Specifically, as observed in the interaction matrix in Figure \ref{fig:interaction-matrix-1}, there is a clear main diagonal, indicating that on average, the intended taste for which the music was generated is recognized. This recognition is more apparent for sweet and bitter music, while sour music is often perceived as bitter, and salty music is frequently associated with sweetness. This aligns with previous discussions about the biases present in the dataset used for fine-tuning. As studied by \cite{wang_turn_2016}, our results show a strict correlation between positive emotions and sweetness and negative feelings with bitterness, confirming that anger and disgust were less used in the ratings, a known fact studied by \cite{mohn_pom_2011}.
These findings are further corroborated by the factor analysis. The factor loadings Table \ref{tab:factor-loadings} highlights that the first factor is dominated by negative adjectives, bitterness, and sourness, with a notable inverse correlation with happiness. In contrast, the second factor is almost exclusively dominated by sweetness, which resonates with warmth and happiness but also with sadness, demonstrating that positive valence can be perceived even in sad music \citep{kawakami_frontiers_2013, sachs_fhn_2015}. The third factor represents temperature, indicating that negative emotions and sour and salty flavors align with cold sensations, while warmth and happiness align in the opposite direction \citep{spence_temperature-based_2020}. The fourth factor combines salty, happiness, surprise, warmth, and sourness. The first, second and fourth factors, when considered in terms of emotional aspects, clearly characterize valence, with positive (factors 2 and 4) and negative (factor 1) dimensions. Temperature appears to be separate from other dimensions, aside from minor, non-significant correlations, suggesting its use as an indicator of perceived arousal from the stimulus.
Furthermore, looking at Figure \ref{fig:interaction-matrix-2}, the prompt ``sour'' showed a higher average response, possibly due to a greater presence of negative scales or confusion with bitterness. The interaction matrix reveals that bitter music is often rated as sad and independent of temperature, while sour music encompasses more negative sensations and is most associated with disgust, a rarely used adjective in musical contexts, as observed in \cite{heike_psychology_2016}.

\section{Conclusions}\label{sec:conclusions}
In this study, we investigated the potential of a fine-tuned generative model to induce synesthetic taste perceptions through music, focusing on the intricate correlations between music, emotions, and taste. The findings revealed that music could indeed evoke gustatory sensations, with positive valence emotions closely aligning with positive valence tastes. Temperature also emerged as a significant factor in these correlations, suggesting a complex interplay between sensory modalities. The results, supported by rigorous ANOVA and factor analysis, underscore the model's capability to bridge sensory modalities, providing valuable insights into the emotional and perceptual connections between sound and taste.
Despite these promising results, the study faced several limitations that must be acknowledged. The sample size was limited to 111 participants, predominantly from the same geographical region and of similar age, which may affect the generalizability of the findings. This homogeneity in the sample could potentially skew the results, limiting their applicability to a broader population. Additionally, while the adjectives used in the study allowed up to a certain degree of freedom in evaluations, they fell short of covering certain aspects necessary to fully encompass Russell's circumplex model of emotions. This gap suggests that the emotional dimensions explored in the study might not capture the full spectrum of human emotional experience. Furthermore, participants were not presented with preparatory stimuli to align their emotional and perceptual states, a factor known to influence perception, as studied by \cite{taylor_psychology_2014} and \cite{rentfrow_jpsp_2011}. This oversight could have introduced variability in the participants' responses, potentially impacting the study's outcomes.

The insights gained from this study have significant implications for various applications. One potential impact is in aiding individuals affected by autism spectrum disorder, by inducing emotional responses through music and food. By harnessing the emotional power of music, it may be possible to facilitate communication and emotional expression in individuals who face difficulties in conventional interactions. Additionally, the concept of ``sonic seasoning'' could be further developed to enhance culinary experiences by aligning music with taste to influence perception and enjoyment. This innovative approach could revolutionize the way we experience food, adding a new dimension to culinary arts and hospitality.
Looking ahead, future research should focus on addressing the limitations identified in this study. Developing a more comprehensive dataset that better represents the diversity of sensory experiences would enhance the accuracy and applicability of the model.
Developing a more sophisticated model could improve the accuracy and depth of synesthetic inductions, enabling more refined applications. Integrating additional modalities, as suggested in \cite{spanio_aixia_2024}, may further enhance results through emotional mediation. The improved performance of the fine-tuned model underscores multimodal AI’s potential to bridge sensory domains, emphasizing the need for well-curated datasets to support innovative cross-modal applications.

\bibliography{spanio}  

\end{document}